\documentclass[twocolumn,prl,aps]{revtex4}
\usepackage{array}
\usepackage{amsmath}
\usepackage{graphicx}
\usepackage{bbm}
\usepackage{xcolor}

\def\ket#1{|#1 \rangle}
\def\bracket#1#2{\langle #1|#2 \rangle}

\begin{document}

\title{Universal state orthogonalizer and qubit generator}

\author{Antonio S. Coelho$^{1}$, Luca S. Costanzo$^{1,2}$, Alessandro Zavatta$^{1,2}$, Catherine Hughes$^{3}$, M. S. Kim$^{3}$, \&  Marco Bellini$^{1,2}$}

\affiliation{$^{1}$Istituto Nazionale di Ottica (INO-CNR), L.go E. Fermi 6, 50125 Florence, Italy\\
$^{2}$LENS and Department of Physics, University of Firenze, 50019 Sesto Fiorentino, Florence, Italy\\
$^{3}$QOLS, The Blackett Laboratory, Imperial College London, SW7 2AZ, UK}

\bigskip
\begin{abstract}
The superposition principle is at the heart of quantum mechanics and at the root of many paradoxes arising when trying to extend its predictions to our everyday world. Schr\"odinger's cat \cite{Sch} is the prototype of such paradoxes and here, in contrast to many others, we choose to investigate it from the operational point of view. We experimentally demonstrate a universal strategy for producing an unambiguously distinguishable type of superposition, that of an arbitrary pure state and its orthogonal. It relies on only a limited amount of information about the input state to first generate its orthogonal one. Then, a simple change in the experimental parameters is used to produce arbitrary superpositions of the mutually orthogonal states. Constituting a sort of Schr\"odinger's black box, able to turn a whole zoo of input states into coherent superpositions, our scheme can produce arbitrary continuous-variable optical qubits, which may prove practical for implementing quantum technologies and measurement tasks.

\end{abstract}

\maketitle

In the original Schr\"odinger's cat paradox, an atom decaying into its ground state $|0\rangle_a$ causes a poisonous gas to be released, meaning the poison operation $\hat{O}_P$ is performed. If the atom instead remains in its excited state $|1\rangle_a$, the gas is not released and no operation is performed, as represented by the identity operator $\mathbbm{1}$. When the atom is at its half-life point, it sits in a superposition of excited and ground states and the total operation $|0\rangle_a\hat{O}_P+|1\rangle_a\hat{\mathbbm{1}}$ applies. The main ingredient of the story is not the cat, but rather the superposition of the two operations. Once it is realized, whatever creature is in the Schr\"odinger's box, it will be put in a superposition of the ``alive'' and ``dead'' orthogonal states. In the quantum mechanical context, states $\psi$ and $\psi_\perp$ are said to be orthogonal when the overlap between the two state vectors is zero, {\it i.e.}, $\langle\psi|\psi_\perp\rangle=0$. 

In the following we introduce and experimentally demonstrate a universal operation which can turn any initial state into its orthogonal. We then show how to simply modify this tool to generate a quantum superposition of the two. 

The importance of quantum superposition states for enhancing the power of information processing has been widely investigated in the last decades, and a complete toolbox of theoretical machinery and experimental techniques has been developed in this rapidly evolving field. The quantum bit, or {\it qubit}, is the basic element in the quantum regime. Contrary to what happens in classical computers, where a bit can either assume the \textbf{0} or \textbf{1} values, quantum mechanics allows one to encode information also in the generic coherent superposition $A \ket{0} + B \ket{1}$ (with $|A|^2+|B|^2=1$), of the two orthogonal state vectors $\ket{0}$ and $\ket{1}$.
Out of an arbitrary initial state vector $\ket{\psi}$ a generic qubit can be obtained by first producing its orthogonal state $\ket{\psi_{\perp}}$, and then implementing their coherent superposition 
\begin{equation}
\ket{\Psi}=A \ket{\psi} + B \ket{\psi_{\perp}}.
\label{eq:qubit}
\end{equation}
As for single-photon qubit states, continuous-variable (CV) ones are possible candidates for realizing arbitrary qubits following this procedure. While CV qubits have usually been represented by superpositions of pairs of opposite-phase coherent states $\ket{\alpha}$  and $\ket{-\alpha}$ \cite{Milburn,Yurke,Brune,Vlastakis2013}, the above scheme allows one to use any initial basis state $\ket{\psi}$.

However, just like it is impossible to perfectly and deterministically clone or amplify a quantum state \cite{Wootters82,Caves82}, the realization of a perfect universal NOT gate, which would turn any arbitrary input state $\ket{\psi}$ into $\ket{\psi_{\perp}}$, is not allowed by quantum mechanics \cite{Buzek99}, without prior information. 
Despite this fundamental limitation, it has been recently shown by Vanner et al. \cite{vanner} that a perfect orthogonalizer can be in principle realized even if only some very limited preliminary information about the input state is available. Given any arbitrary operator $\hat{C}$, it is sufficient to know its mean value $\langle \hat{C} \rangle$ on the input state $\ket{\psi}$, to build the general orthogonalizer 
\begin{equation}
\hat{O}_{C}\equiv(\hat{C}-\langle \hat{C} \rangle \mathbbm{1}).
\label{eq:ortho}
\end{equation} 
It is straightforward to see that when $\hat{O}_{C}$ is applied to the input state, the result is an orthogonal state $\ket{\psi_{\perp}}$, such that $\bracket{\psi}{\psi_{\perp}}=0$. Although the operator $\hat{C}$ can be in principle arbitrary, the above procedure cannot be applied if the input states are among its eigenstates because its success probability drops to zero. An orthogonalizer of this kind was demonstrated very recently in two-dimensional systems by Jezek et al. \cite{jezek}. Once the orthogonalizer is in operation, any superposition of the original state and its orthogonalized state can be realized by the superposition of $\hat{O}_{C}$ and the identity $\mathbbm{1}$.
\begin{equation}
\ket{\Psi}\propto [\hat{C}+(c-\langle \hat{C} \rangle )\mathbbm{1}] \ket{\psi},
\label{eq:qube}
\end{equation}
where $c$ is a coefficient determined by the superposition weight.

\begin{figure}[h]
\includegraphics[width=1\linewidth]{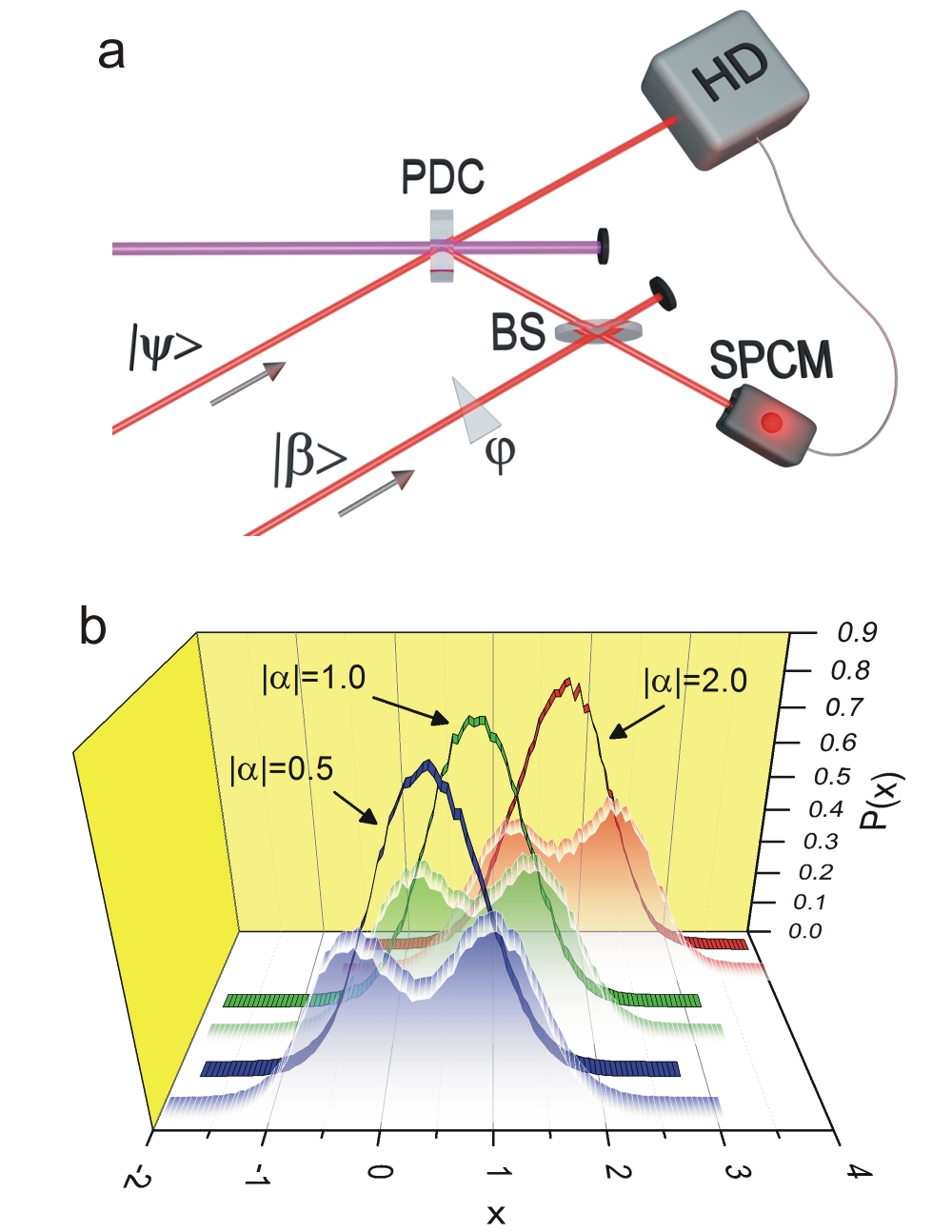}
\caption{\textbf{State orthogonalizer and CV qubit generator based on the photon creation operator.} a. Conceptual experimental scheme of the orthogonalizer and CV qubit generator based on photon addition by heralded stimulated PDC. A click in the single-photon-counting module (SPCM) normally heralds a single photon addition to the input $\ket{\Psi}$ state. However, if the PDC idler mode is mixed with a coherent state $\ket{\beta}$ on a beam-splitter (BS) prior to detection, a superposition of the photon creation operator and the identity with adjustable weights and phases can be obtained. In the actual experiments we used coherent states $\ket{\alpha}$ as the input states and the operator superposition was implemented by using polarization modes (see Methods for details). HD is a time-domain homodyne detector triggered by SPCM clicks.  b. Measured $x$ quadrature distributions (marginals of the Wigner function) for the input coherent states and for the corresponding results of the orthogonalization procedure with $\alpha= 0.5, 1.0, 2.0$.}
    \label{fig:setup}
\end{figure}

Here we present the first experimental implementation of the generic orthogonalization procedure of Eq.\ref{eq:ortho}, and use it as a basis to build arbitrary CV superposition states as those of Eq.\ref{eq:qubit}.
A particularly simple and interesting case in this scheme is obtained when $\hat{C}\equiv\hat{a}^{\dag}$, the bosonic creation operator, which has no eigenstates and can thus be safely applied independently of the arbitrary state at the input. 
Here, one just needs to know the mean value of $\hat{a}^{\dag}$ on the particular input state to construct the $(\hat{a}^{\dag}-\langle \hat{a}^{\dag} \rangle \mathbbm{1} )$ orthogonalizer or a general qubit as prescribed by Eq.\ref{eq:qube}. Both these operations can be experimentally implemented by extending some of the tools recently developed in our group. In particular, the photon creation operator can be conditionally realized by means of stimulated parametric down-conversion (PDC) in a nonlinear crystal seeded by the optical input state in the signal mode \cite{Zavatta2004,zavatta05}. The coherent superposition of this operation and the identity can be realized by mixing the (herald) idler PDC mode with a coherent light field on an unbalanced beam-splitter that erases the information about the origin of a click in the heralding single-photon detector at one of its outputs. By simply controlling both the relative phase between the input and the coherent state impinging on the beam-splitter, and the reflectivity of the latter, different superpositions of $\hat{a}^{\dag}$ and $\mathbbm{1}$ can be obtained, in particular those corresponding to the orthogonalizer and to arbitrary CV superposition states.  A simplified scheme of the experimental setup is shown in Fig.\ref{fig:setup}a (also see Methods). 

Similar techniques, involving phase-space displacement on the herald mode of conditional state generation, have been recently used for quantum state engineering up to two photons \cite{bimbard10,Yukawa2013}, and for the generation of optical CV qubits made of superpositions of squeezed vacuum and squeezed single-photon states \cite{jonas10}.

We tested the concept presented above by using coherent states $\ket{\alpha}=\hat{D}(\alpha)|0\rangle$ as the input, where $\hat{D}(\alpha)$ is the displacement operator \cite{Glauber}. In this special case, the general orthogonalizer operator becomes
\begin{equation}
\hat{O}_{a^{\dag}}\equiv(\hat{a}^{\dag}-\alpha^* \mathbbm{1} ),
\label{eq:orthoalpha}
\end{equation}
 and it is easy to see that, when applied to $ \ket{\alpha} $, this results in the displaced Fock state $\hat{D}(\alpha)\ket{1}$, which is clearly orthogonal to $\ket{\alpha}$. The repeated application of the $ \hat{O}_{a^{\dagger}} $ operator to $ \ket{\alpha} $ then produces a set of mutually orthogonal states that can be a base for qudit encoding in a larger Hilbert space (see Methods).
Fig.\ref{fig:setup}b illustrates the result of the application of the orthogonalizer to coherent states of different initial amplitudes. The $x$ quadrature distributions clearly show that the Wigner functions of the orthogonal states are differently displaced versions of a single-photon Wigner function.
\begin{figure}[h]
\includegraphics[width=1\linewidth]{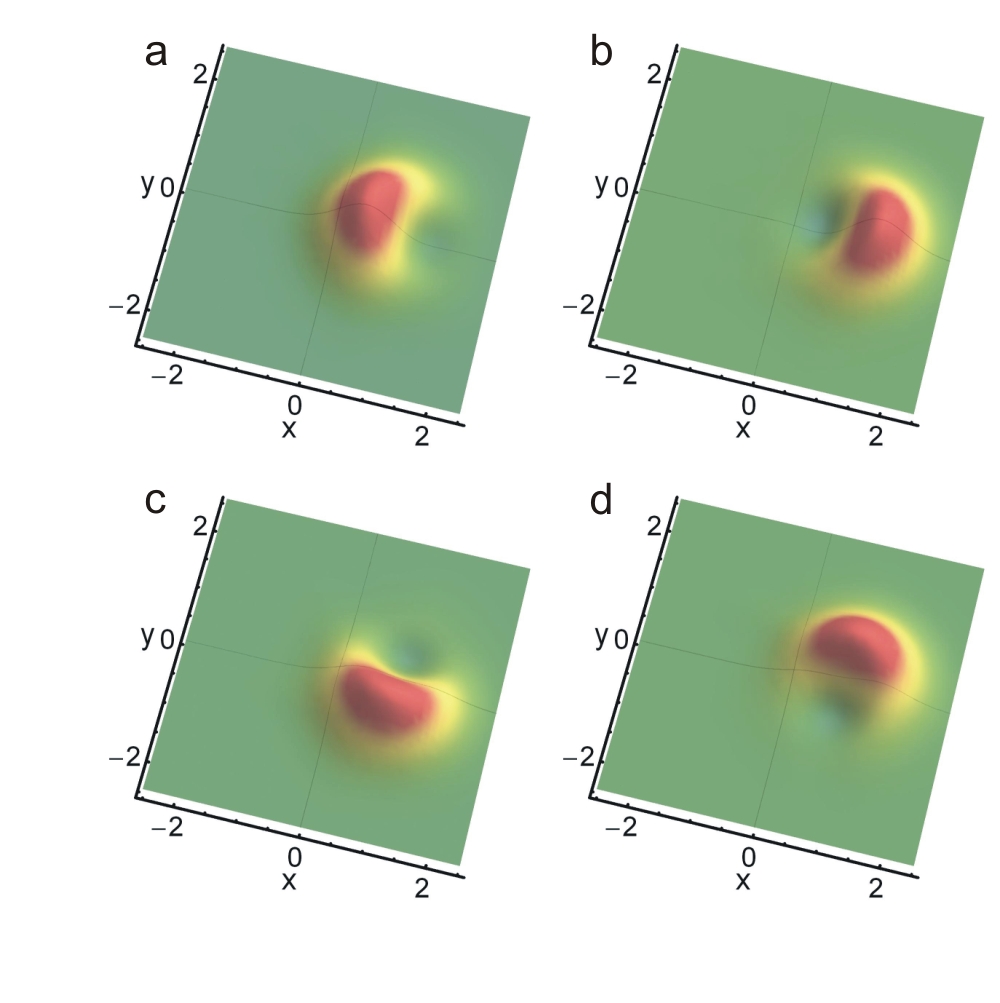}
\caption{\textbf{Wigner functions of arbitrary CV qubit states.} Wigner functions for different, balanced, superpositions of states $\ket{\alpha}$ and $\ket{\alpha_{\perp}}$ with $\alpha=1$, as reconstructed without correcting for the limited detection efficiency. a) and b) correspond to states $1/\sqrt{2}(\ket{\alpha} \pm \ket{\alpha_{\perp}})$, c) and d) to states $1/\sqrt{2}(\ket{\alpha} \pm i \ket{\alpha_{\perp}})$, respectively.} 
    \label{fig:ortho}
\end{figure}

Although there are other, probably simpler, ways to produce the same final states as those obtained in this specific case \cite{Lvovsky2002}, we stress that our approach has a more general scope than this, and the final states reduce to this particular case only when the photon creation operator is used in combination with input coherent states. Our goal is entirely different, since we are not aiming to produce a particular quantum state, but rather to demonstrate a universal scheme for producing orthogonal and CV qubit states starting from arbitrary, meaning either non-classical or classical, inputs. 

In fact, just a simple adjustment of the parameters in the beam-splitter placed in the idler mode allows one to produce various CV qubit states. In Fig.\ref{fig:ortho} we show the measured Wigner functions for different equal-weight superposition states of an input coherent state with $ \alpha\approx 1$ and of its orthogonal. In the different plots the phase of the resulting CV qubit is simply varied by properly controlling the relative phase between the input and the displacement coherent states and the reflectivity of the idler beam-splitter.

In the above examples, the photon creation operation $\hat{a}^{\dag}$ was used as the operator entering the general orthogonalizer recipe of Eq.\ref{eq:ortho}. However, the mean photon number of the state may often be more easily obtained, rather than the mean value of the creation operator. In such cases one may insert the number operator $\hat{n}\equiv\hat{a}^{\dag}\hat{a}$ and the mean photon number $\bar{n}$ in Eq.\ref{eq:ortho}, which thus becomes   
\begin{equation}
\hat{O}_{n}\equiv(\hat{n}-\bar{n} \mathbbm{1}).
\label{eq:ortho_n}
\end{equation} 

\begin{figure}[h]
\includegraphics[width=1\linewidth]{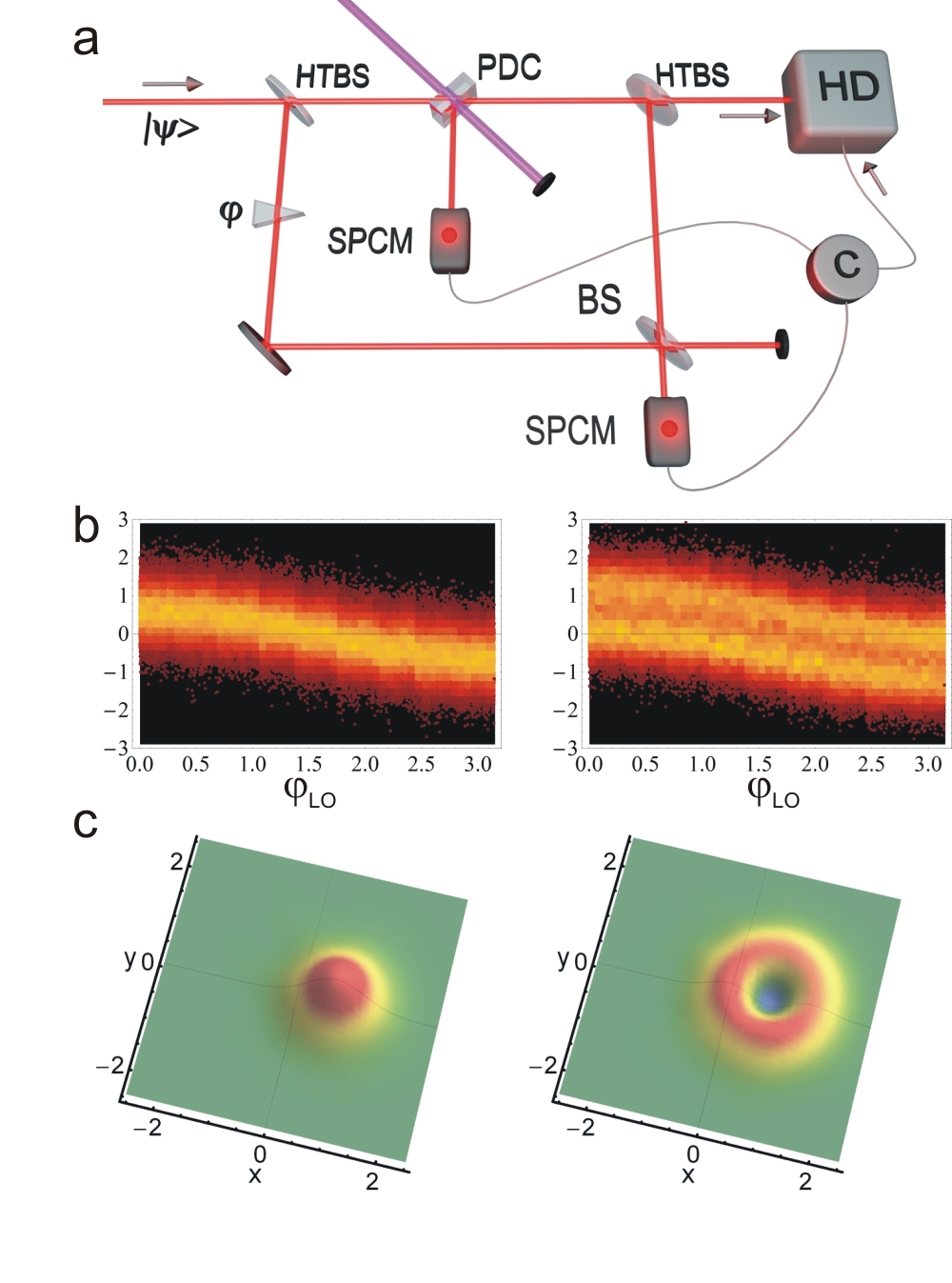}
\caption{\textbf{State orthogonalizer based on the photon number operator.} a) Conceptual experimental scheme for the orthogonalizer based on the photon number operator. HTBS are high-transmittivity beam-splitters, C is a coincidence logic circuit. b) Experimental homodyne detection traces for the original input coherent state $\ket{\alpha}$  (left panel) and for the state obtained by the orthogonalization procedure (right panel). The input amplitude was $\alpha\approx 1$, and 10 values of the local oscillator phase $\varphi_{LO}$ in the $[0, \pi)$ interval were used. c) Wigner functions of the input coherent state and of its orthogonal, as reconstructed from the homodyne data without correcting for the limited detection efficiency.}
    \label{fig:setupn}
\end{figure}

In order to verify the effectiveness and generality of the proposed approach, we also put this scheme to an experimental test with the setup of Fig.\ref{fig:setupn}a, which is similar to the one first developed for testing the bosonic commutation relation \cite{Zavatta2009} (see Methods). Such a setup conditionally produces the arbitrary superposition of operators
\begin{equation}
(A\hat{a}^{\dag}\hat{a}+B\hat{a}\hat{a}^{\dag}),
\label{eq:ortho_aa}
\end{equation} 
which is seen to be proportional to $(\hat{n}+\frac{B}{A+B}\mathbbm{1})$ using the bosonic commutation relation. Therefore, the generic orthogonal state to one of mean photon number $\bar{n}$ can be straightforwardly implemented by adjusting the setup of Fig.\ref{fig:setupn}a so that $\frac{B}{A+B}=-\bar{n}$. This approach, where the full orthogonalizing operation is performed using two operators, can also be viewed as a generalization of Eq.\ref{eq:ortho}: $\hat{O}_C=\hat{C}_1-\frac{\langle\hat{C}_1\rangle}{\langle\hat{C}_2\rangle}\hat{C}_2$. It is seen that, when used in combination to input coherent states $\ket{\alpha}$, this scheme results in the same orthogonal state as for the previous example. Fig.\ref{fig:setupn}b and c show the measured quadrature distributions and reconstructed Wigner functions for such input and orthogonal output states, as obtained for $ \alpha\approx 1$.

In conclusion, we have shown the first experimental application of a universal orthogonalization procedure to arbitrary CV optical states. Relying on a very limited amount of preliminary information about the input states, we verified the effectiveness and generality of this powerful technique through the illustrative examples of the photon creation and photon number operators. Simple modifications in the experimental parameters also allowed us to produce CV qubits based on the superposition of an arbitrary input state $\ket{\psi}$ and its orthogonal $\ket{\psi_{\perp}}$, in close analogy to a Schr\"odinger's box able to turn any input creature into a superposition state. Our general scheme can be applied to various physical systems including phononic states of ions in a trap and nanomechanical oscillators \cite{vanner}. Besides its immediate implications in the context of quantum information processing, the production and control of the most distinguishable type of superposition in terms of unambiguous discrimination, that of a pure state and its orthogonal state, may also represent a first step toward the test of quantum-to-classical transition models \cite{review1,review2,GRW,Penrose,Leggett,Zurek}. 

\bigskip

\section*{Appendix}

\subsubsection{Experimental setup}
We use a mode-locked Ti:sapphire laser emitting a train of 786 nm picosecond pulses at a repetition rate of 80 MHz for producing the input coherent states $\ket{\alpha}$ and, after frequency doubling to 393 nm, as the pump for degenerate, non-collinear, parametric down-conversion in a 3-mm long $\beta$-barium borate crystal. We obtain stimulated PDC by synchronously seeding the crystal with the coherent state pulses along the signal mode. Heralded photon addition in such a mode is conditioned on a single photon detection by an avalanche photodiode (SPCM) placed in the idler mode after narrow spectral and spatial filters (see \cite{Zavatta2004,zavatta05}).

For the orthogonalizer and CV qubit realization based on the photon creation operator and schematically depicted in Fig.\ref{fig:setup}a, we need to make the clicks of the SPCM due to photon addition indistinguishable from those coming from an ancillary coherent state $\ket{\beta}$. Differently from the conceptual scheme shown in the figure, we don't combine the modes on a beam splitter, but rather make use of the polarization degrees of freedom. The ancillary coherent state pulses $\ket{\beta}$ are originated from the same laser pulse train and injected along the PDC idler channel with an horizontal polarization. Since we exploit type-I parametric down-conversion with an horizontally-polarized pump, such an idler injection does not contribute to stimulated emission, nor it interferes with the vertically-polarized seed $\ket{\alpha}$ coherent pulses injected along the signal mode. By placing a polarizer, slightly rotated from the vertical position, in front of the SPCM we can adjust the relative weight of the photon addition and identity operations in the operator superposition. Their relative phase is controlled and actively locked by translating a mirror mounted on a piezoelectric transducer in the path of the $\ket{\beta}$ pulses.

The experimental setup used for the orthogonalizer based on the photon number operator and schematically depicted in Fig.\ref{fig:setupn}a is based on that first used for the implementation of the commutator operator \cite{Zavatta2009}.  It relies on the coherent superposition of two inverted sequences of photon creation and annihilation operators by means of a beam-splitter (BS) that mixes the heralding modes of the two photon-subtraction operations. A coincidence between the clicks from two single-photon detectors, one at the output of the beam-splitter, and the other on the herald mode of the photon addition stage placed between the two subtraction ones, certifies that a sequence of photon addition and subtraction has certainly taken place, but leaves their order unknown. By adjusting the beam-splitter reflectivity and the relative phase between the two subtraction herald modes, a generic superposition of the two sequences like in Eq.\ref{eq:ortho_aa} can thus be realized. The relative weights between the two inverted sequences of operations is then properly adjusted according to the mean photon number of the input states.

The conditionally generated output states are finally analyzed by means of a high-frequency, time-domain homodyne detector \cite{zavatta02:josab} triggered by single SPCM clicks in the case of the scheme utilizing the photon creation operator or coincidences between the clicks from two SPCM detectors (one for photon addition and one for a delocalized photon subtraction) in the photon number operator case.

\subsubsection{Theory considerations}

A set of orthogonal states may be constructed by repeated application of the operator $\hat{O}_{a^{\dag}}\equiv(\hat{a}^{\dag}-\alpha^* \mathbbm{1} )$ to $ \ket{\alpha} $. Furthermore, the result will be the same for $\hat{O}_{a^{\dag}}\hat{O}_{a^{\dag}}|\alpha\rangle$, and on to states of form $\hat{O}^n_{a^{\dag}}$ for $n\in\mathbbm{Z^+}$. Thus with this operation we can construct a set of mutually orthogonal states.

We now consider the theoretical implementation of the scheme in Fig.\ref{fig:setup}a, which generates the operation of $\hat{O}_{a^{\dag}}$. The PDC operation is to add a photon to $|\psi\rangle$, heralded by a photon generated in the second mode \cite{MSKrev}. Including the beam-splitting operation between this second mode and the coherent state $|\beta\rangle$, the total operation is represented by $\hat{B}(|1\rangle\hat{a}^{\dagger}+|0\rangle\mathbbm{1})|\beta\rangle|\psi\rangle$, where the input $|\psi\rangle$ is in the same mode as the bosonic operators $\hat{a}$ and $\hat{a}^{\dagger}$.
Here, $\hat{B}$ is the beam-splitter operator \cite{MSKrev}, which acts on coherent states mixed with the vacuum as $\hat{B}|0\rangle|\beta\rangle = |{-}r\beta\rangle|t\beta\rangle$, where $r$ and $t$ are the reflectivity and transmittivity of the beam-splitter. The final step in this method is to project the expression onto $|1\rangle|0\rangle$, yielding
\begin{equation}\label{eq:ortho-operator}
t\hat{a}^{\dagger}-r\beta\mathbbm{1},
\end{equation}
up to an overall global phase which is omitted to better see the similarity of the result to operator $\hat{O}_{a^{\dag}}$. Using the unitary transformation by $\hat{D}(\alpha)$ of $\hat{a}^{\dagger}$ \cite{MSKrev} and Eq. \ref{eq:orthoalpha}, it is straightforward to see
\begin{align}
\hat{O}^n_{a^{\dagger}}|\alpha\rangle = \hat{O}^n_{a^{\dagger}}\hat{D}(\alpha)|0\rangle &= \hat{D}(\alpha)(\hat{a}^{\dagger})^n|0\rangle = \sqrt{n!}\hat{D}(\alpha)|n\rangle.
\end{align}
Thus $\hat{O}^m_{a^{\dagger}}|\alpha\rangle \perp \hat{O}^n_{a^{\dagger}}|\alpha\rangle$ for any $m\neq n$.

Such an expression may also be produced for the photon number operator $\hat{n}=\hat{a}^{\dagger}\hat{a}$, where ideally the operator is $\hat{O}_n=\hat{n}-\bar{n}\mathbbm{1}$, with $\bar{n}$ the mean photon number. This approach, described in Fig.\ref{fig:setupn}, begins with the expression $\hat{B}(\mathrm{e}^{i\phi}\hat{a}^{\dagger}\hat{a}|1\rangle|0\rangle+\hat{a}\hat{a}^{\dagger}|0\rangle|1\rangle),$ where this time it is noted that $\hat{B}|1\rangle|0\rangle = t|1\rangle|0\rangle+r|0\rangle|1\rangle$. Following the measurement again of $|1\rangle|0\rangle$, the resulting expression, up to a phase, is
\begin{equation}
\hat{n}-\frac{r}{t-r}\mathbbm{1},
\end{equation}
where $\frac{r}{t-r}$ can be set to equal $\bar{n}$ with the appropriate adjustment of the beam-splitter.

\section{Acknowledgments}A.Z., L.S.C., and M.B. acknowledge the support of the EU under the ERA-NET CHIST-ERA project QSCALE, and of the MIUR under the contract FIRB RBFR10M3SB. M.S.K. and C.H. thank the UK EPSRC for their financial support and Peter Knight and Kimin Park for discussions. A.S.C. acknowledges the support of the CNPq-Brazil.

\section{Correspondence} Correspondence and requests for materials should be addressed to M.B. (email: bellini@ino.it) and/or M.S.K. (email: m.kim@imperial.ac.uk).

\end{document}